\documentclass[aps,preprint,showpacs,eqsecnum]{revtex4}
\usepackage{graphicx}
\usepackage{amsmath}
\usepackage{amssymb}
\begin{document}
\title{Solutions for correlations along the coexistence curve
and at the critical point of a kagom\'e lattice gas with
three-particle interactions}
\author{J. H. Barry and K. A. Muttalib}
\email[Corresponding author:~]{muttalib@phys.ufl.edu} \affiliation
{Department of Physics, University of Florida, P.O. Box 118440,
Gainesville, FL 32611-8440}
\author{T. Tanaka}
\email[Professor Emeritus:~] {ttanaka@sannet.ne.jp} \affiliation
{Department of Physics and Astronomy, Ohio University, Athens, OH
45701}

\begin{abstract}

We consider a two-dimensional (d=2) kagom\'e lattice gas model
with attractive three-particle interactions around each triangular
face of the kagom\'e lattice. Exact solutions are obtained for
multiparticle correlations along the liquid and vapor branches of
the coexistence curve and at criticality. The correlation
solutions are also determined along the continuation of the
curvilinear diameter of the coexistence region into the disordered
fluid region. The method generates a linear algebraic system of
correlation identities with coefficients dependent only upon the
interaction parameter. Using a priori knowledge of pertinent
solutions for the density and elementary triplet correlation, one
finds a closed and linearly independent set of correlation
identities defined upon a spatially compact nine-site cluster of
the kagom\'e lattice. Resulting exact solution curves of the
correlations are plotted and discussed as functions of the
temperature, and are compared with corresponding results in a
traditional kagom\'e lattice gas having nearest-neighbor pair
interactions. An example of application for the multiparticle
correlations is demonstrated in cavitation theory.


\end{abstract}
\maketitle

\section{Introduction}

Our fundamental understanding of anomalous thermodynamic and
transport behaviors in a many-body cooperative system stems in
large measure from concomitant knowledge and applications of its
thermal equilibrium correlations. The familiar representations of
macroscopic observables in terms of their underlying correlations,
e.g., specific heat and magnetic susceptibility as energy and
magnetization fluctuations, respectively, Kubo formulas,
temperature-dependent Green's functions, fluctuation-dissipation
theorems, and so forth, constitute many of the most instructive
basic relationships and useful formulations in statistical physics
and bestride virtually all areas of theoretical investigation in
phase transitions, critical and multi-critical phenomena. Indeed,
the impetus for the modern unifying interpretation of critical
phenomena was the recognition of the essential role that the
anomalously long-ranged spatial correlations played near a
critical point resulting in scaling theories and the
renormalization group approach towards problems in phase
transitions and particle physics. Since spatial correlation
functions are structured using thermal expectation values of
\textit{products} of local variables, they clearly offer a more
detailed description than thermodynamic for the order and symmetry
in the system, and a precise presentation of the correlation
solutions becomes highly desirable.

Besides examining the asymptotically large distance behavior of
correlation functions, it is also useful to obtain solutions for
more spatially compact, short-distance type correlations
\cite{ghosh}. These smaller scale correlations have varied
applications, at criticality and otherwise. More particularly,
\textit{Ising-type models} have special appeal since, in select
cases, their localized correlations \cite{montroll} can be
calculated exactly (any Hamiltonian having a finite density of
finitely discrete commuting local variables can be cast as a
standard or generalized Ising model). Actually, the archetypical
two-dimensional (d=2) Ising model \cite{mccoy} magnet with
nearest-neighbor pair interactions and in zero magnetic field is
the only realistic microscopic model of cooperative phenomena for
which many correlation solutions have been obtained exactly. Ising
models \cite{ising} are employed not only to represent certain
kinds of highly anisotropic magnetic crystals but also, e.g., as
lattice models for liquids, alloys, adsorbed monolayers,
equilibrium polymerization, for biological and chemical systems,
and in field theories of elementary particles (lattice gauge
theories describing the quark structure of hadrons). Applications
for localized correlations in Ising-type models occur in the
analysis of local equilibrium properties in the vicinity of
isolated defects \cite{fisher1}, in the theory of both transport
coefficients \cite{mahan} and thermodynamic response functions
\cite{fisher2}, in investigations of inelastic neutron scattering
\cite{allan} , percolation phenomena \cite{essam}, and in many
other problems including topical connections between entanglements
and spin correlations in quantum information theory \cite{popp},
and a present example (Section VI) in cavitation theory
\cite{brennen}.

Due to severe mathematical complexities, few analytically rigorous
results are known for models having multiparticle interactions
\cite{note1}. The lack of exact solutions for correlations in the
models is an incentive for the present theoretical investigations.
In the present paper, one obtains exact solutions for
multiparticle correlations along the coexistence curve and at
criticality of a kagom\'e lattice gas with localized
three-particle interactions. The phase diagrams for condensation
of the lattice gas were determined previously \cite{barry},
specifically, the chemical potential and the density versus
temperature. The theory was based upon work by Wu \cite{wu1}, Wu
and Wu \cite{wu2}, and Lin and Chen \cite{lin}, who established
that the partition function of a generalized (three-parameter)
kagom\'e Ising model having pair and triplet interactions and
magnetic field is equivalent, aside from known pre-factors, to the
partition function of a standard (two-parameter) honeycomb Ising
model with pair interactions and field. Their theoretical
developments incorporated a symmetric eight-vertex model on the
honeycomb lattice in a mediating role. Later, the same result was
obtained by Wu \cite{wu3} using a direct mapping without a
weak-graph transformation.

Foreknowledge of the above phase diagrams (chemical potential and
density vs temperature) is vital supplementary information in the
present theory, as is securing the solution for the elementary
triplet correlation via appropriate logarithmic differentiation of
the grand canonical partition function of the lattice gas. A
linear algebraic system of correlation identities is generated
with coefficients dependent solely upon the interaction parameter
of the model when evaluated along the coexistence curve. Using the
known pertinent solutions for the density $\langle n_0\rangle$ and
the elementary triplet correlation $\langle n_0n_1n_2\rangle$ as a
priori information, one succeeds in finding a closed and linearly
independent set of correlation identities defined upon a spatially
compact nine-site cluster of the kagom\'e lattice. Resulting exact
solution curves of the correlations are plotted and discussed as
functions of the temperature, and are compared with the
corresponding results in a traditional half-filled kagom\'e
lattice gas having nearest-neighbor pair interactions. To our
knowledge, these are the first examples, certainly away from
criticality, of exact solutions for multiparticle correlations in
any planar lattice-statistical model with multiparticle
interactions. Finally, the solutions for the multiparticle
correlations are applied to cavitation theory in the condensation
of the lattice gas.

The paper is organized as follows. Section II presents the
triplet-interaction kagom\'e lattice gas model, and reviews its
grand partition function and phase boundary curve. Section III
reviews the liquid-vapor coexistence curve of the model, and
calculates the relevant elementary triplet correlation. Section IV
derives a basic generating equation for developing a linear
algebraic system of correlation identities. Taking advantage of
the supplemental information in Section III, Section V solves a
set of the identities to secure solutions for multiparticle
correlations along the coexistence curve and at criticality.
Section VI provides an example of application for the
multiparticle correlations in cavitation theory. Lastly, Section
VII is a summary and discussion.

\section{Partition function and phase
boundary curve of a kagom\'e lattice gas with three particle
interactions}

Consider a lattice gas of $N$ atoms upon the kagom\'e lattice
(Figure 1) of $\mathcal{N}$ sites with the (dimensionless)
Hamiltonian
\begin{equation}
-\beta {\mathcal{H}} = K_3\sum_{<i,j,k>}n_i n_j n_k,
\end{equation}
where $\beta\equiv 1/k_BT$, with $k_B$ being the Boltzmann
constant and $T$ the absolute temperature, $K_3=\beta \epsilon_3$
with $\epsilon_3 > 0$ being the strength parameter of the
short-range attractive triplet interaction, the sum is taken over
all elementary triangles, and the idempotent site occupation
numbers are defined as
\begin{equation}
n_l=\begin{cases}1 & \textrm{ site $l$ occupied}, \\
0 & \textrm{ site $l$ empty}.\end{cases}
\end{equation}
In (2.1), an infinitely-strong (hard core) repulsive pair
potential has also been tacitly assumed for atoms on the {\it
same} site, thereby preventing multiple occupancy of any site as
reflected in the occupation numbers (2.2).

In the usual context of the grand canonical ensemble, one
introduces
\begin{equation}
{\mathrm{H}}\equiv {\mathcal{H}} -\mu N
\end{equation}
where $\mu$ is the chemical potential with $N$ being the conjugate
total number of particles
\begin{equation}
N=\sum_in_i\;\; .
\end{equation}
Using (2.1), (2.3) and (2.4), the grand canonical partition
function $\Xi(\mu,{\mathcal{N}},T)$ is given by
\begin{equation}
\Xi(\mu,{\mathcal{N}},T)=\sum_{\{n_i\}}e^{-\beta{\mathrm{H}}}
=\sum_{\{n_i\}}e^{\beta\mu\sum_in_i+K_3\sum_{<i,j,k>}n_in_jn_k},
\end{equation}
where the summation symbol $\{n_i\}$ represents the set of total
$\mathcal{N}$ occupation numbers. Aside from known pre-factors,
the grand canonical partition function (2.5) on the kagom\'e
lattice can be transformed into the magnetic canonical partition
function
\begin{equation}
Z^*(L^*,K^*)=\sum_{\{\mu_i\}}e^{-\beta\mathcal{H}^*_{hc}}=\sum_{\{\mu_i\}}
e^{L^*\sum_i\mu_i+K^*\sum_{<i,j>}\mu_i\mu_j},
\end{equation}
$\mu_{\ell}=\pm1$, $\ell=1, \cdots, \mathcal{N}^*$, of a standard
$S=1/2$ Ising model ferromagnet upon the associated honeycomb
lattice (Figure 2) with (dimensionless) Hamiltonian
$-\beta\mathcal{H}_{hc}^*$ having a (dimensionless) external
magnetic field $L^*$ and (dimensionless) nearest-neighbor pair
interaction parameter $K^*>0$, and where the summation symbol
$\{\mu_i\}$ represents the set of total
$\mathcal{N}^*(=2\mathcal{N}/3)$ Ising variables. Specifically
\cite{barry},
\begin{subequations}
\begin{eqnarray}
\Xi(\mu,{\mathcal{N}},T)&=&\sum_{\{n_i\}}e^{\beta\mu\sum_in_i+K_3\sum_{<i,j,k>}n_in_jn_k}\cr
&=&e^{\frac{3}{4}(\frac{K_3}{6}+\beta\mu)\mathcal{N}^*}(a^*/2\cosh
L^*)^{\mathcal{N}^*}(\cosh
K^*)^{-\frac{3}{2}\mathcal{N}^*}Z^*(L^*,K^*),
\end{eqnarray}
with the parameters
\begin{eqnarray}
a^* &=& e^{-\frac{1}{8}(K_3+6\beta\mu+12\ln 2)}\;\;
\frac{1+(e^{K_3}-1)(1+e^{-\beta\mu})^{-3/2}}{[1+(1+e^{-\beta\mu})^{-1/2}]^{3/2}},
\\
L^* &=& \frac{1}{2}\ln (e^{K_3}-1)-\frac{3}{4}\ln
(1+e^{-\beta\mu}),
\\
K^* &=& \frac{1}{4}\ln (1+e^{-\beta\mu}),
\end{eqnarray}
\end{subequations}
and where the positivity $K^*>0$ (ferromagnetic) is manifest in
(2.7d).

It is well known \cite{griffiths} that a necessary and sufficient
condition for the existence of a phase transition in the $S=1/2$
\textit{ferromagnetic} ($K^*>0$) honeycomb Ising model is the
joint condition $L^*=0$ and $K^*\geq K_c$, where the critical
value $K^*_c=\frac{1}{2}\ln(2+\sqrt{3})=0.65847\cdots$. Imposing
this joint condition of the associated Ising model upon the
current calculations enables an exact solution to be found for the
phase boundary curve of the triplet interaction kagom\'e lattice
gas model. In particular, the zero field condition $(L^*=0)$ is
realized by setting (2.7c) to zero which implies that the chemical
potential $\mu$ is prescribed by the relation
\begin{equation}
e^{-\beta\mu}=(e^{K_3}-1)^{2/3}-1 \;\;\; \textrm{at}\;\;\; L^*=0.
\end{equation}
Substituting (2.8) into (2.7d) gives
\begin{equation}
K^*=\frac{1}{6}\ln (e^{K_3}-1) \;\;\; \textrm{at}\;\;\; L^*=0,
\end{equation}
relating the interaction parameters $K^*$, $K_3$ whenever the
magnetic field parameter $L^*=0$. Using (2.9), the manifest
positivity $K^*>0$ from (2.7d) is therefore equivalent, at
$L^*=0$, to $\ln 2<K_3<\infty$. In the next section, the
restricted range $\ln 2<K_3<\infty$ also assures that the solution
found for the average particle number density $\rho$ does not
exceed unity (fully occupied lattice gas). For the remaining range
$0 \le K_3 < \ln 2$, the present model (2.1) doesn't admit phase
transitions or critical behavior since the necessary zero-field
($L^*=0$) condition (2.8) cannot be satisfied by any real chemical
potential $\mu$.

\textit{At criticality}, the aforementioned literature value
$K^*_c=\frac{1}{2}\ln(2+\sqrt{3})$ is substituted into (2.9)
yielding
\begin{equation}
K_{3c}=\ln[(2+\sqrt{3})^3+1]=3.96992\cdots.
\end{equation}
As comparison, for a traditional kagom\'e lattice gas with
attractive nearest-neighbor \textit{pair} interactions, the
corresponding critical value $K_{2c}$ is known to be
\cite{mccoy,naya}
\begin{equation}
K_{2c}=\ln(3+2\sqrt{3})=1.86626\cdots,
\end{equation}
which is nearly $50\%$ smaller than the critical value (2.10). The
critical value $K_{3c}$ in (2.10) is used to locate the critical
point in the relation (2.8), yielding the \textit{liquid-vapor
phase boundary curve} of the triplet interaction kagom\'e lattice
gas (Figure 3a). Explicitly, one directly obtains \cite{barry}
\begin{equation}
\mu/\epsilon_3=-K^{-1}_3\ln[(e^{K_3}-1)^{2/3}-1], \;\;\; 0 \leq
K_{3c}/K_3\leq 1,
\end{equation}
with $\mu/\epsilon_3$ being a reduced chemical potential and
$K_{3c}/K_3(=T/T_c)$ a reduced temperature where
$K_{3c}(\equiv\epsilon_3/k_BT_c)=\ln[(2+\sqrt{3})^3+1]=3.96992\cdots$
[(2.10)]. The curvilinear phase boundary begins at zero
temperature with $\mu/\epsilon_3=-2/3$ and ends (analytically) at
a critical point whose coordinates are $K_{3c}/K_3=1$,
$\mu/\epsilon_3=\mu_c/\epsilon_3=-0.64469\cdots$. At zero
temperature, the phase boundary curve $\mu/\epsilon_3$ vs $T/T_c$
has a zero slope in accordance with the Clausius-Clapeyron
equation and third law of thermodynamics. Otherwise, its slope is
\textit{positive} which is more discernible at temperature closely
below the critical temperature. As comparison, Figure 3b shows the
corresponding phase diagram for the condensation of a conventional
kagom\'e lattice gas with attractive nearest-neighbor pair
interactions.

\section{Exact solutions for $\langle n_0\rangle$, $\langle n_0n_1n_2\rangle$ along coexistence
curve of kagom\'e lattice gas with three-particle interactions}

The coexistence surface (and boundary) of the lattice-gas system
is the prominent portion of its thermal equation of state surface.
More particularly, the boundary (edge) of the coexistence surface
contains the critical point and encloses the liquid-vapor
coexistence region (heterogeneous mixture). The projective mapping
of the boundary onto the chemical potential-temperature ($\mu-T$)
plane marks the previous liquid-vapor phase boundary curve (2.12),
and a similar projective viewing in density-temperature ($\rho-T$)
space indicates the \textit{liquid-vapor coexistence curve} of the
lattice gas. The latter phase diagram ($\rho^{coex}_{l,v}$ vs
$T/T_c$) for the condensation of the triplet-interaction kagom\'e
lattice gas has already been obtained \cite{barry}. A brief review
of the results is now presented for later use.

The solutions for the density $\rho(=\langle n_0\rangle)$ and the
elementary triplet correlation $\langle n_0n_1n_2\rangle$ can be
determined along the liquid-vapor coexistence curve and at
criticality by logarithmic differentiations of the grand partition
function $\Xi(\mu, \mathcal{N},T)$ [(2.7)] with respect to
$\beta\mu$ and $K_3$, respectively, and then letting
$L^*\rightarrow 0$ for the full range of condensation temperatures
$0\leq K_{3c}/K_3\leq1$. Specifically, the exact solution for the
average number density $\langle n_0\rangle(=\rho)$ is given, as
$L^*\rightarrow 0$, by \cite{barry}
\begin{equation}
\rho= \begin{cases} 1-\frac{1}{4} [1-(e^{K_3}-1)^{-2/3}][1+\langle
\mu_0\mu_1\rangle_{L^*=0}\mp 2\langle \mu\rangle_S], & 0 \leq
\frac{K_{3c}}{K_3}\leq 1, \\ 1-\frac{1}{4}
[1-(e^{K_3}-1)^{-2/3}][1+\langle \mu_0\mu_1\rangle_{L^*=0}], & 1 <
\frac{K_{3c}}{K_3}\leq \frac{K_{3c}}{\ln 2}=5.72739\cdots
,\end{cases}
\end{equation}
where $\langle \mu_0\mu_1\rangle_{L^*=0}$, $\langle \mu\rangle_S$
are the nearest-neighbor pair correlation and spontaneous
magnetization, respectively, of the previous $S=1/2$ honeycomb
Ising model ferromagnet. In the top of eq.~(3.1), the $-(+)$
algebraic sign corresponds to the liquid(vapor) branch of the
coexistence curve. The restricted range $\ln 2 < K_3 < \infty$ was
required earlier to assure that the Ising interaction parameter
$K^*>0$ (ferromagnetic) and leads, using (2.10), to the finite
$K_{3c}/K_3-$ range of the temperatures in the bottom of
eq.~(3.1). The terminating value $K_{3c}/\ln 2=5.72739\cdots$ also
guarantees that $\rho$ does not exceed $\rho_{max}=1$
corresponding to the fully occupied (``close packed'') lattice
gas. The bottom expression of (3.1) is the continuation of the
curvilinear diameter of the coexistence region in the top
expression of (3.1) beyond the coexistence surface. In (3.1), the
exact solutions for $\langle \mu_0\mu_1\rangle_{L^*=0}$ and
$\langle \mu\rangle_S$ are known to be \cite{baxter,naya}
\begin{subequations}
\begin{eqnarray}
\langle\mu_0\mu_1\rangle_{L^*=0} &=& \frac{2}{3}[\coth 2K^*+\gamma
K_1(\kappa)],
\\
\langle \mu\rangle_S &=& \begin{cases}(1-\kappa^2)^{1/8}, \;\;\; 0
\leq \frac{K^*_{c}}{K^*}\leq 1, \\ 0, \;\;\; 1 <
\frac{K^*_{c}}{K^*}< \infty \end{cases}
\end{eqnarray}
\end{subequations}
with $K_1(\kappa)$ being the complete elliptic integral of the
first kind
\begin{subequations}
\begin{equation}
K_1(\kappa)=\int_0^{\pi/2}(1-\kappa^2\sin^2\theta)^{-1/2}d\theta,
\end{equation}
and where
\begin{eqnarray}
\kappa^2 &=& 16z^3(1+z^3)(1-z)^{-3}(1-z^2)^{-3},
\\
\gamma &=& (1-z^4)(z^2-4z+1)/\pi|1-z^2|(1-z)^4,
\\
z &=& e^{-2K^*}.
\end{eqnarray}
\end{subequations}

The expressions (3.1) can be written purely in the natural $K_3$
notation of the lattice gas by substituting the interaction
parameter relation (2.9) into (3.2) and (3.3). Then, substituting
the resulting forms into the top expression of (3.1) yields the
sought exact solution for the liquid-vapor coexistence curve which
is plotted in Figure 4a. Additionally, the composition of the
density $\rho$ at points within the two-phase coexistence region
in Figure 4a can be determined by the application of the
thermodynamic \textit{lever rule} \cite{callen}. The phase diagram
of Figure 4a exhibits an asymmetric rounded shape with a
curvilinear diameter, contrasting the familiar symmetric rounded
shape and constant rectilinear diameter for a conventional
kagom\'e lattice gas with attractive nearest-neighbor pair
interactions (see Figure 4b). In the global comparison (i.e.,
``over-laying'')of Figure 4a and Figure 4b, the values of the
liquid-vapor branches (and curvilinear diameter) in Figure 4a
exceed the values of the liquid-vapor branches (and rectilinear
diameter) in Figure 4b at all corresponding finite reduced
temperatures.

The curvilinear diameter of the coexistence region in Figure 4a is
the solid curve which begins at zero temperature with $\rho=1/2$,
exhibits a positive slope which is more pronounced at temperatures
closely below the critical temperature, and ends at the critical
point (solid circle) whose coordinates are $K_{3c}/K_3=1$,
$\rho=\rho_c=0.58931\cdots$ (see (3.7a)). Using the bottom
expression of (3.1), the curvilinear diameter is continued beyond
the coexistence region as the solid sigmoidal curve, eventually
ending at the point (solid square) with coordinates
$K_{3c}/K_3=K_{3c}/\ln 2=5.72739\cdots$, $\rho=\rho_{max}=1$.

To obtain the exact solution for the triplet correlation $\langle
n_0n_1n_2\rangle$ along the coexistence curve, one first computes
the logarithmic derivative of $\Xi(\mu,\mathcal{N},T)$ with
respect to the triplet-interaction parameter $K_3$. Specifically,
(2.7a) gives
\begin{eqnarray}
\ln\Xi(\mu,{\mathcal{N}},T)&=&
\frac{3}{4}(\frac{K_3}{6}+\beta\mu)\mathcal{N}^*+
\mathcal{N}^*(\ln a^*-\ln\cosh L^*-\ln 2)\nonumber \\ &-&
\frac{3}{2}\mathcal{N}^*\ln\cosh K^* + \ln Z^*(L^*,K^*),
\end{eqnarray}
where the total number of kagom\'e lattice sites
$\mathcal{N}=\frac{3}{2}\mathcal{N}^*$ with $\mathcal{N}^*$ being,
as stated previously, the total number of lattice sites of the
associated honeycomb lattice (see Figure 2). Also, the total
number of elementary triangles $\mathcal{N}_{\Delta}$ of the
kagom\'e lattice equals $\mathcal{N}^*$, again seen in Figure 2.
Using (2.6), (2.7) and (3.4), one obtains
\begin{subequations}
\begin{eqnarray}
\mathcal{N}_{\Delta}\langle n_0n_1n_2\rangle &=& \frac{\partial
\ln \Xi(\mu,{\mathcal{N}},T)}{\partial K_3}\cr &=&
\frac{\mathcal{N}^*}{8} +\mathcal{N}^*\left[\frac{\partial\ln
a^*}{\partial K_3}-(\tanh L^*)\frac{\partial L^*}{\partial
K_3}\right]\cr &-& \frac{3}{2}\mathcal{N}^* (\tanh K^*)
\frac{\partial K^*}{\partial K_3}+ \frac{\partial \ln
Z^*}{\partial L^*}\frac{\partial L^*}{\partial K_3}+
\frac{\partial \ln Z^*}{\partial K^*}\frac{\partial K^*}{\partial
K_3}
\end{eqnarray}
where
\begin{eqnarray}
\frac{\partial \ln Z^*}{\partial L^*} &=& \mathcal{N}^* \langle
\mu_i\rangle,
\\
\frac{\partial L^*}{\partial K_3} &=& \frac{1}{2}
\frac{e^{K_3}}{e^{K_3}-1}\; ,
\\
\frac{\partial K^*}{\partial K_3} &=& 0,
\\
\frac{\partial \ln a^*}{\partial K_3} &=& -\frac{1}{8}+
\frac{e^{K_3}}{(e^{-\beta\mu}+1)^{3/2}+e^{K_3}-1}\; .
\end{eqnarray}
\end{subequations}
Hence, as $L^*\rightarrow 0$, (3.5) yields
\begin{eqnarray}
\langle n_0n_1n_2\rangle &=&  \frac{1}{2}
\frac{e^{K_3}}{e^{K_3}-1}(1\pm\langle \mu\rangle_S), \;\;\; \ln 2 < K_3 < \infty \nonumber \\
&=& \left\{\begin{array}{ll}
\frac{1}{2}\frac{e^{K_3}}{e^{K_3}-1}(1\pm\langle \mu\rangle_S),
\;\;\; 0 \leq \frac{K_{3c}}{K_3}\leq 1, \\ \frac{1}{2}
\frac{e^{K_3}}{e^{K_3}-1},  \;\;\; 1 < \frac{K_{3c}}{K_3}\leq
\frac{K_{3c}}{\ln 2}=5.72739\cdots ,\end{array} \right.
\end{eqnarray}
having replaced $\langle\mu_i\rangle$ by $\pm \langle\mu\rangle_s$
in (3.5b) due to `spontaneous symmetry breaking', having
substituted the zero-field ($L^*=0$) condition (2.8) into (3.5e),
used the identification $\mathcal{N}_{\Delta}=\mathcal{N}^*$, and
having partitioned, as earlier, the (dimensionless) inverse
temperature range $\ln 2 < K_3 < \infty$ into  (reduced)
temperature intervals below and above criticality in (3.6). Using
the known result (3.2b) for the spontaneous magnetization $\langle
\mu\rangle_S$, and substituting the interaction parameter relation
(2.9) into (3.2b) and (3.3b,d), the exact solution in the upper
expression of (3.6) for the elementary triplet correlation
$\langle n_0n_1n_2\rangle$ is obtained along the coexistence curve
and at criticality as a function of the (reduced) temperature
$K_{3c}/K_3$, and is plotted in Figure 5a. Also shown in Figure
5a, the lower expression of (3.6) is the exact solution for
$\langle n_0n_1n_2\rangle$ along the continuation of the
curvilinear diameter (solid curve) beyond the coexistence surface,
where the trajectory terminates at the point (solid square) with
coordinates $K_{3c}/K_3=5.72739\cdots$, $\langle
n_0n_1n_2\rangle=1$. The corresponding results for $\langle
n_0n_1n_2\rangle$ in a kagom\'e lattice gas with attractive
nearest-neighbor pair interactions are shown in Figure 5b where,
as a function of temperature, the curvilinear diameter and its
extension into the disordered fluid region exhibit a sigmoidal
shape and monotonically decreasing behavior contrasting the
results shown in Figure 5a for the three-particle interactions.

\textit{At criticality}, the honeycomb Ising ferromagnet has
values $\langle \mu\rangle_S=0$ and \cite{syozi} $\langle
\mu_0\mu_1\rangle_c=4\sqrt{3}/9$. These critical values along with
the critical value (2.10) $e^{K_{3c}}=(2+\sqrt{3})^3+1$ are
substituted into (3.1) and (3.6) yielding, respectively,
\begin{subequations}
\begin{eqnarray}
\rho_c &=& \frac{1}{3}(\frac{7}{2}-\sqrt{3})=0.58931\cdots,
\\
\langle n_0n_1n_2\rangle_c&=&
\frac{1}{2}(\frac{27+15\sqrt{3}}{26+15\sqrt{3}})=0.50961\cdots,
\end{eqnarray}
\end{subequations}
as critical values in the triplet-interaction kagom\'e lattice
gas. For a traditional kagom\'e lattice gas with attractive
nearest-neighbor \textit{pair} interactions, the corresponding
values are
\begin{subequations}
\begin{eqnarray}
\rho_c &=& \frac{1}{2}=0.50000\cdots,
\\
\langle n_0n_1n_2\rangle_c &=&
\frac{1}{16}(3+2\sqrt{3})=0.40400\cdots.
\end{eqnarray}
\end{subequations}

As seen, the one-parameter ($\epsilon_3$) model Hamiltonian (2.1)
exhibits a \textit{single} critical point on its thermal equation
of state surface $\mu=\mu(\rho,T)$ (in suitable units), with
critical coordinates $\mu_c/\epsilon_3=-0.64469\cdots$,
$\rho_c=0.58931\cdots$, $k_BT_c/\epsilon_3\equiv
K^{-1}_{3c}=0.25189\cdots$. The coexistence surface, its boundary
coexistence curve and the critical point comprise the prominent
portion of the above equation of state surface. The liquid and
vapor branches of the coexistence curve identify at the critical
point, and all lattice-gas thermal averages in the current studies
are evaluated along both branches and at criticality.

Some comments are warranted on the nature of the mathematical
singularities in the phase diagram of Figure 4a. The critical
behaviors of the coexistence curve and the curvilinear diameter
are underlaid by the known critical behaviors of the honeycomb
Ising thermal averages $\langle \mu\rangle_s$ and $\langle
\mu_0\mu_1\rangle_{L^*=0}$ in (3.1). Letting $\rho_l, \rho_v$
denote the particle number density along the liquid and vapor
branches, respectively, of the coexistence curve, the top
expression in (3.1) immediately reveals that the \textit{ordering
parameter} $\rho_l-\rho_v$ (length of vertical ``tie-line''
spanning the coexistence region) realizes its critical behavior
purely from $\langle \mu\rangle_s$, whereas the
\textit{curvilinear diameter} $\frac{1}{2}(\rho_l+\rho_v)$
(arithmetic mean of $\rho_l$ and $\rho_v$) realizes its critical
behavior solely from $\langle \mu_0\mu_1\rangle_{L^*=0}$. Hence,
the relation (3.2b) for the Ising spontaneous magnetization
$\langle \mu\rangle_s$ leads to the vanishing of the order
parameter $\rho_l-\rho_v$ at the critical point with the
Ising-type critical exponent $1/8$ (algebraic branch point
singularity). Similarly, relation (3.2a) for the Ising
nearest-neighbor pair correlation $\langle
\mu_0\mu_1\rangle_{L^*=0}$ leads to the result that the
curvilinear diameter $\frac{1}{2}(\rho_l+\rho_v)$ and its analytic
continuation into the disordered fluid region (lower expression of
(3.1)) possess a weak Ising energy-type singularity \cite{mermin}
$\epsilon \ln\epsilon$ at the critical point, where $\epsilon > 0$
is a small fractional deviation of the temperature from its
critical value. In contrast, note that the constant rectilinear
diameter in a conventional $d=2$ lattice gas (Figure 4b) is
analytic at criticality. One also recognizes that the curvilinear
diameter of the solution curve for the elementary triplet
correlation $\langle n_0n_1n_2\rangle$ in (3.6) and Figure 5a is
analytic at the critical point. In the above mathematical
arguments, one uses (2.9), (2.10) and the previous literature
value $K^*_{c}=\frac{1}{2}\ln (2+\sqrt{3})$ to establish an
\textit{exact scaling relation}
\begin{equation}
\epsilon^*=\frac{1}{6}\left(\frac{e^{K_{3c}}}{e^{K_{3c}}-1}\right)
\frac{K_{3c}}{K^*_c}\;\epsilon =(1.02416 \cdots )\epsilon
\end{equation}
between the smallness parameters $\epsilon^*$ and $\epsilon$. The
scaling relation (3.9) affords a direct proof that the
singularities in the associated Ising model lead to the
\textit{same} nature of singularities in the lattice gas phase
diagram in (3.1) and Figure 4a, viz., $\epsilon^{*1/8}\rightarrow
\epsilon^{1/8}$, $\epsilon^*\ln \epsilon^* \rightarrow \epsilon\ln
\epsilon$, where, neglecting second-order small quantities,
$\epsilon^*=(K^*-K^*_c)/K^*_c$, $\epsilon=(K_3-K_{3c})/K_{3c}$.

In the present section, we emphasized that the solutions for
$\langle n_0\rangle$ and $\langle n_0n_1n_2\rangle$ were
determined along the coexistence curve via logarithmic
differentiations of the grand partition function
$\Xi(\mu,{\mathcal{N}},T)$ with respect to $\beta\mu$ and $K_3$,
respectively, and then letting $L^*\rightarrow 0$. To secure
solutions for additional correlations, a different method is
needed. In the subsequent sections, one develops and solves, as
$L^*\rightarrow 0$, a linear algebraic system of correlation
identities whose coefficients depend solely upon the interaction
parameter $K_3$. The a priori knowledge of $\langle n_0\rangle$
and $\langle n_0n_1n_2\rangle$ will be instrumental in the quest
for closure and linear independence within the linear algebraic
system of identities.

\section{Basic generating equation for correlation identities}

The class of correlation identities currently considered is a set
of linear algebraic equations with coefficients dependent only
upon  the interaction parameter $K_3$ \cite{fisher}. To develop
such identities systematically, one proceeds to derive their basic
generating equation.

Let [g] be any function of the lattice-gas variables $n_1, n_2,
\cdots , n_{\mathcal{N}-1}$ (\textit{excluding} $n_0$, the origin
site variable in Figure 1). Letting
$H^{\prime}\equiv\mathcal{H}^{\prime}-\mu N^{\prime}$,
$\sum^{\prime}_{\{n_i\}}$ denote a \textit{restricted} energy form
and summation operation, respectively, which \textit{exclude}
$n_0$, one can construct the grand canonical thermal average
$\langle n_0[g]\rangle$ as
\begin{eqnarray}
\Xi\cdot\langle n_0[g]\rangle &=& \sum_{\{n_i\}}\;\;
n_0[g]e^{-\beta H} \cr &=& {\sum_{\{n_i\}}}^{\prime}\;\; [g]
e^{-\beta H^{\prime}}\sum_{n_0} n_0e^{n_0[\beta\mu +
K_3(n_1n_2+n_3n_4)]}\cr &=& \sum_{\{n_i\}}\;\; [g] e^{-\beta H}
\left[\frac{\sum_{n_0} n_0e^{n_0[\beta\mu +
K_3(n_1n_2+n_3n_4)]}}{\sum_{n_0} e^{n_0[\beta\mu +
K_3(n_1n_2+n_3n_4)]}}\right]
\end{eqnarray}
yielding
\begin{equation}
\langle n_0[g]\rangle=\langle\frac{e^{\beta\mu +
K_3(n_1n_2+n_3n_4)}}{1+e^{\beta\mu +
K_3(n_1n_2+n_3n_4)}}[g]\rangle, \;\;\; n_0 \nsubseteq [g],
\end{equation}
having lastly used the standard definition of grand canonical
thermal average  initiating (4.1). In deriving (4.1), the ``split,
rearrange, then reconstitute'' procedures are justified since all
lattice-gas variables commute. To further develop (4.2), one
writes
\begin{equation}
\frac{e^{\beta\mu + K_3(n_1n_2+n_4n_4)}}{1+e^{\beta\mu +
K_3(n_1n_2+n_3n_4)}}= A+B(n_1n_2+n_3n_4)+Cn_1n_2n_3n_4,
\end{equation}
where the expansion as a \textit{finite} algebraic series in the
lattice-gas product variables $n_1n_2$, $n_3n_4$ reflects their
idempotent nature [$(n_pn_q)^2=n_pn_q$]. The coefficients $A$,
$B$, $C$ are determined by considering the following realizations
of the product variables $n_1n_2,n_3n_4$ in (4.3):
\begin{subequations}
\begin{itemize}
\item[(i)] $n_1n_2=n_3n_4=0$, yielding
\begin{equation}
A=\frac{e^{\beta\mu}}{1+ e^{\beta\mu}} =\frac{1}{e^{-\beta\mu}+1},
\end{equation}
\item[(ii)] $n_1n_2=0$, $n_3n_4=1$, yielding
\begin{equation}
A+B=\frac{e^{\beta\mu+K_3}}{1+ e^{\beta\mu+K_3}}
=\frac{1}{e^{-\beta\mu-K_3}+1},
\end{equation}
\item[(iii)] $n_1n_2=n_3n_4=1$, yielding
\begin{equation}
A+2B+C=\frac{e^{\beta\mu+2K_3}}{1+ e^{\beta\mu+2K_3}}
=\frac{1}{e^{-\beta\mu-2K_3}+1}.
\end{equation}
\end{itemize}
\end{subequations}
The three linear algebraic inhomogeneous equations (4.4 a-c) in
the three unknowns $A$, $B$, $C$ are patently linearly-independent
and directly give the solutions
\begin{subequations}
\begin{eqnarray}
A &=& \frac{1}{e^{-\beta\mu}+1},
\\
B &=& \frac{1}{e^{-\beta\mu-K_3}+1}-\frac{1}{e^{-\beta\mu}+1},
\\
C &=&\frac{1}{e^{-\beta\mu-2K_3}+1}-\frac{2}{e^{-\beta\mu-K_3}+1}
 + \frac{1}{e^{-\beta\mu}+1}.
\end{eqnarray}
\end{subequations}
For adoption along the coexistence curve, one evaluates the
coefficients (4.5) at $L^*=0$. Substituting the (2.8) expression
into (4.5), one obtains, as $L^*\rightarrow 0$,
\begin{subequations}
\begin{eqnarray}
A &=& (e^{K_3}-1)^{-2/3},
\\
B &=& e^{K_3}[(e^{K_3}-1)^{2/3}+e^{K_3}-1]^{-1}-
(e^{K_3}-1)^{-2/3},
\\
C &=&  e^{2K_3}[(e^{K_3}-1)^{2/3}+e^{2K_3}-1]^{-1}  \nonumber
\\
 &-& 2e^{K_3}[(e^{K_3}-1)^{2/3}+e^{K_3}-1]^{-1}+ (e^{K_3}-1)^{-2/3}.
\end{eqnarray}
\end{subequations}

Returning to (4.2) and substituting (4.3), the \textit{basic
generating equation} for multiparticle correlation identities is
thus given by
\begin{equation}
\langle n_0[g]\rangle=A\langle [g]\rangle + B \langle
(n_1n_2+n_3n_4) [g]\rangle +C \langle n_1n_2n_3n_4[g]\rangle,
\;\;\; n_0 \nsubseteq [g],
\end{equation}
where the coefficients $A$, $B$, $C$ are given by (4.6) as
$L^*\rightarrow 0$. The linear algebraic system of correlation
identities generated by (4.7) will be employed in the next section
to determine the solutions for various multiparticle correlations
along the coexistence curve of the triplet-interaction kagom\'e
lattice gas.

\section{Exact solutions for correlations along the coexistence curve of
kagom\'e lattice gas with three-particle interactions. }

Solutions will now be determined for correlations along the
coexistence curve of the present lattice gas model. More
particularly, exact solutions are found for the nearest-neighbor
pair and various multiparticle correlations defined upon the
select \textit{nine-site} cluster shown in Figure 6. In section
III, pertinent exact solutions were obtained for $\langle
0\rangle$ and $\langle 012\rangle$, where, for notational
simplicity, only the numeric site labels in Figure 6 are written
inside the thermal average symbols. As seen shortly, the a priori
knowledge of $\langle 0\rangle$ and $\langle 012\rangle$ is vital
supplementary information within the system of correlation
identities.

First consider the \textit{five-site} (``bow-tie'') cluster
$0,1,2,3,4$ (see Figure 6) and the five generators $\langle
0\rangle$, $\langle 01\rangle$, $\langle 012\rangle$, $\langle
0123\rangle$, $\langle 01234\rangle$. Employing the basic
generating equation (4.7), one directly obtains the following
correlation identities for $\langle 0\rangle$ and $\langle
012\rangle$, respectively,
\begin{subequations}
\begin{eqnarray}
\langle 0\rangle &=& A+B\langle 12+34 \rangle +C \langle 1234
\rangle,
\\
\langle 012 \rangle &=& A\langle 12 \rangle+B\langle 12+1234
\rangle +C\langle 1234 \rangle,
\end{eqnarray}
\end{subequations}
having used the idempotent property $n^2_j=n_j$. The identity
(5.1a) is initially the \textit{only} inhomogeneous equation in
the infinite system of identities. However, since $\langle
0\rangle$ and $\langle 012\rangle$ are already known along the
coexistence curve, one rearranges (5.1) into the standard form of
two linear algebraic inhomogeneous equations in the two unknowns
$\langle 01\rangle$ and $\langle 1234\rangle$:
\begin{subequations}
\begin{eqnarray}
2B\langle 01 \rangle + C\langle 1234 \rangle &=& \langle 0 \rangle
-A,
\\
(A+B)\langle 01 \rangle + (B+C)\langle 1234 \rangle &=& \langle
012 \rangle,
\end{eqnarray}
\end{subequations}
where the symmetry of the kagom\'e lattice has been recognized in
the equating of geometrically-equivalent pair correlations in
(5.2). Since the LHS coefficient matrix in (5.2) has a
non-vanishing determinant, equations (5.2) are linearly
independent and hence determine the solutions for the
nearest-neighbor pair $\langle 01\rangle$ and quartet $\langle
1234\rangle$ correlations along the coexistence curve.

Again employing the basic generating equation (4.7), the three
remaining generators $\langle 01\rangle$,  $\langle 0123\rangle$
and $\langle 01234\rangle$ yield, respectively, the identities
\begin{subequations}
\begin{eqnarray}
\langle 01 \rangle &=& A\langle 1 \rangle +B\langle 12+134 \rangle
+C \langle 1234 \rangle ,
\\
\langle 0123 \rangle &=& (A+B)\langle 123 \rangle +(B+C)\langle
1234 \rangle ,
\\
\langle 01234 \rangle &=& (A+2B+C)\langle 1234 \rangle .
\end{eqnarray}
\end{subequations}
The relevant exact solutions for the thermal averages $\langle
0\rangle$, $\langle 01\rangle$, $\langle 1234\rangle$ appearing in
(5.3) have already been obtained. Thus, using lattice symmetry
($\langle 1\rangle=\langle 0\rangle$, $\langle 12\rangle=\langle
01\rangle$, $\langle 134\rangle=\langle 123\rangle$), the
identity(5.3a) directly determines the triplet correlation
$\langle 123\rangle$, identity (5.3b) in turn directly establishes
the quartet generator $\langle 0123\rangle$ since the RHS is
known, and identity (5.3c) similarly secures the quintet generator
$\langle 01234\rangle$ since the RHS is known. To review, exact
solutions are now known along the coexistence curve for the
density $\langle 0\rangle$ and \textit{six} correlations defined
upon the \textit{five-site} (``bow tie'') cluster $0,1,2,3,4$ in
Figure 6, specifically,
\begin{equation}
\langle 0 \rangle, \langle 01 \rangle, \langle 012 \rangle,
\langle 123 \rangle, \langle 0123 \rangle, \langle 1234
\rangle,\langle 01234 \rangle.
\end{equation}
One next considers the \textit{seven-site} cluster $0,1,2,3,4,5,6$
(see Figure 6) and the six generators $<03456>$, $<0456>$,
$<01256>$, $\langle 012356\rangle$, $\langle 012456\rangle$,
$\langle 0123456\rangle$. Using the basic generating equation
(4.7), the latter generators yield, respectively, the identities
\begin{subequations}
\begin{eqnarray}
\langle 03456 \rangle &=& (A+B)\langle 3456 \rangle +(B+C)\langle
123456 \rangle ,
\\
\langle 0456 \rangle &=& A\langle 456 \rangle +B\langle 3456+12456
\rangle +C \langle 123456 \rangle ,
\\
\langle 01256 \rangle &=& (A+B)\langle 1256 \rangle +(B+C)\langle
123456 \rangle,
\\
\langle 012356 \rangle &=& (A+B)\langle 12356 \rangle
+(B+C)\langle 123456 \rangle,
\\
\langle 012456 \rangle &=& (A+B)\langle 12456 \rangle
+(B+C)\langle 123456 \rangle,
\\
\langle 0123456 \rangle &=& (A+2B+C)\langle 123456 \rangle .
\end{eqnarray}
\end{subequations}
Using the lattice symmetry, some correlations in identities (5.5)
on the seven-site cluster are in registry with known correlations
(5.4) on the earlier five-site cluster. Specifically,
\begin{equation}
\langle 03456 \rangle \equiv \langle 01234 \rangle, \langle 3456
\rangle = \langle 0456 \rangle \equiv \langle 0123 \rangle,
\langle 456 \rangle \equiv \langle 012 \rangle.
\end{equation}
The registry listings (5.6) are now used as a priori information
in identities (5.5) thereby reducing the number of unknown
correlations. Hence, in identity (5.5a), all correlations are
known except the sextet correlation $\langle
123456\rangle(=\langle 012356\rangle)$, so the latter is
determined. Similarly, in identity (5.5b), all correlations are
now known except the quintet correlation $\langle 12456\rangle$,
so the latter is obtained. In identity (5.5c), all correlations
are known save the quartet correlation $\langle 1256\rangle$, so
the latter is secured. Continuing the cascade reasoning, the only
unknown correlation in (5.5d) is the quintet correlation $\langle
12356\rangle$ so the latter is found.In identity (5.5e), all
correlations are known save the sextet generator $\langle
012456\rangle$, so the latter is determined. Lastly, in identity
(5.5f), the RHS correlation is known so the LHS septet generator
$\langle 0123456\rangle$ is obtained. In summary, exact solutions
have been found along the coexistence curve for the additional
\textit{six} correlations defined upon the \textit{seven-site}
cluster $0,1,2,3,4,5,6$ in Figure 6:
\begin{equation}
\langle 1256 \rangle, \langle 12356 \rangle, \langle 12456
\rangle, \langle 123456 \rangle, \langle 012456 \rangle, \langle
0123456 \rangle.
\end{equation}
One proceeds to consider the \textit{nine-site} cluster
$0,1,2,3,4,5,6,7,8$ (see Figure 6) and the six generators $\langle
0345678\rangle$, $\langle 012345678\rangle$, $\langle
05678\rangle$, $\langle 045678\rangle$, $\langle 0125678\rangle$,
$\langle 01245678\rangle$. Employing again the basic generating
equation (4.7), the above generators yield, respectively,
\begin{subequations}
\begin{eqnarray}
\langle 0345678 \rangle &=& (A+B)\langle 345678 \rangle
+(B+C)\langle 12345678 \rangle ,
\\
\langle 012345678 \rangle &=& (A+2B+C)\langle 12345678 \rangle,
\\
\langle 05678 \rangle &=& A\langle 5678 \rangle +B\langle
125678+345678 \rangle +C \langle 12345678 \rangle,
\\
\langle 045678 \rangle &=& A\langle 45678 \rangle +B\langle
1245678+345678 \rangle +C \langle 12345678 \rangle,
\\
\langle 0125678 \rangle &=& (A+B)\langle 125678 \rangle
+(B+C)\langle 12345678 \rangle,
\\
\langle 01245678 \rangle &=&  (A+B)\langle 1245678 \rangle
+(B+C)\langle 12345678 \rangle.
\end{eqnarray}
\end{subequations}
Using lattice symmetry, various correlations in (5.8) on the
nine-site cluster are in registry with known correlations (5.7) on
the previous seven-site cluster. Specifically,
\begin{eqnarray}
\langle 0345678 \rangle & \equiv & \langle 0123456 \rangle,
\langle 345678 \rangle\equiv \langle 012456 \rangle, \langle 05678
\rangle\equiv \langle 12356 \rangle, \nonumber \\ \langle 5678
\rangle &\equiv & \langle 1256 \rangle, \langle 045678
\rangle\equiv \langle 123456 \rangle, \langle 45678 \rangle\equiv
\langle 12456 \rangle.
\end{eqnarray}
The registry listings (5.9) are now used as a priori information
in (5.8) thus reducing the number of unknown correlations and
enabling one to elicit further correlation solutions from the
system of identities. In identity (5.8a), all correlations are
known except the octet correlation $\langle 12345678\rangle$ so
the latter is determined. Hence, the RHS of identity (5.8b) is
established thereby yielding the LHS nonuplet generator $\langle
012345678\rangle$. In identity (5.8c), all correlations are known
save the sextet correlation $\langle 125678\rangle$ so the latter
is obtained. In identity (5.8d), all correlations are known save
the septet correlation $\langle 1245678\rangle$ thus the latter is
secured. In identity (5.8e), both RHS correlations are known so
the LHS septet generator $\langle 0125678\rangle$ is found.
Lastly, in identity (5.8f), both RHS correlations are known, hence
the LHS octet generator $\langle 01245678\rangle$ is determined.
In summary, exact solutions have been obtained along the
coexistence curve for the additional \textit{six} correlations
defined upon the \textit{nine-site} cluster $0,1,2,3,4,5,6,7,8$ in
Figure 6:
\begin{equation}
\langle 125678 \rangle, \langle 1245678 \rangle, \langle 01245678
\rangle, \langle 0125678 \rangle, \langle 12345678 \rangle,
\langle 012345678 \rangle.
\end{equation}

To review, one considered correlations defined upon the nine-site
cluster (Figure 6) of the kagom\'e lattice comprising a central
triangle with its three corner-sharing triangles. The collective
contents of (5.4), (5.7) and (5.10) reveal that exact solutions
have been found for the density $\langle 0\rangle$,
nearest-neighbor pair correlation $\langle 01\rangle$, and
\textit{seventeen} multiparticle correlations, all along the
coexistence curve  and at criticality, of the triplet-interaction
kagom\'e lattice gas. Also, the solution for each thermal average
was determined along a finite-length critical isofield ($L^*=0$)
trajectory in the disordered fluid region of $\rho-T$ space. In
this context, illustrative correlations $\langle 01\rangle$,
$\langle 1234\rangle$, $\langle 01234\rangle$, $\langle
012345678\rangle$ are each plotted as functions of the reduced
temperature $x=K_{3c}/K_3 (=T/T_c)$ in Figures 7a,8,9,10,
respectively. Along the coexistence curve (Figure 4a), each
correlation solution exhibits an asymmetric rounded shape with an
infinite slope at the critical point (solid circle). The degree of
asymmetry is associated with the slope of the curvilinear diameter
in the graph, which is more pronounced at temperatures closely
below the critical temperature ($x=1$). One observes that the
slope is \textit{positive} for the elementary triplet and
nearest-neighbor pair correlations in Figures 5a and 7a,
respectively, and \textit{negative} for the quartet, quintet and
nonuplet correlations in Figures 8, 9 and 10, respectively. A
positive (negative) slope of the curvilinear diameter implies that
the correlation solution curve along the upper liquid branch falls
slower (faster) than the corresponding rise of the solution curve
along the lower vapor branch. In each case of negative slope of
the curvilinear diameter (Figures 8, 9, 10), the correlation
solution along the continuation of the curvilinear diameter into
the disordered region shows a rounded \textit{minimum} before
ascending to the common maximum value (solid square) associated
with the fully occupied lattice.

\section{Example of Application}

A simple example employing multiparticle correlations occurs in
cavitation theory \cite{brennen}.  One seeks the probability of an
elementary cavity or void in the fluid system, where an elementary
cavity is construed to be an empty site with all its
nearest-neighbor sites occupied. In a kagom\'e lattice gas, the
joint probability $p_{0,1,1,1,1}$ that the origin site 0 is empty
and its four nearest-neighbor sites $1,2,3,4$ are occupied equals
the thermal average value $\langle(1-n_0)n_1n_2n_3n_4\rangle$.
This probabilistic interpretation of the thermal average can be
directly extracted from the standard definition of average value
in probability theory, as will now be shown.

Let $f(n_0,n_1...,n_4)$ be a function of the lattice-gas discrete
random variables $n_0, n_1,...,n_4$ (see ``bow-tie'' cluster in
Figure~1). Then, the average value $\langle
f(n_0,n_1,...,n_4)\rangle$ is defined by
\begin{equation}
\langle f(n_0,n_1,...n_4)\rangle =\sum_{\{n_0,n_1,...,n_4\}}
p_{n_0,n_1,...,n_4} f(n_0,n_1,...,n_4),\label{eq61}
\end{equation}
where $p_{n_0,n_1,...,n_4}$ is the joint probability of a
\textit{specific} realization of the variables $n_0,n_1,...,n_4$
and the summation symbol $\{n_0,n_1,...,n_4\}$ represents the set
of \textit{all} possible $2^5=32$ realizations of the five lattice
gas variables.

Considering a select function
\begin{equation}
f(n_0,n_1,...,n_4) = (1-n_0) n_1n_2n_3n_4 ,\label{eq62}
\end{equation}
the definition (\ref{eq61}) immediately ``filters out'' the
desired result
\begin{equation}
\langle (1-n_0)n_1n_2n_3n_4 \rangle = p_{0,1,1,1,1} ,\label{eq63}
\end{equation}
since the value of $f(0,1,1,1,1)=1$ and all other values of the
function (\ref{eq62}) vanish in the weighted summation
(\ref{eq61}).  This completes the proof.

It is instructive to mention that taking thermal averages of
select operator functions is a direct and general method for
exactly structuring the elements of reduced statistical density
matrices \cite{note2} in equilibrium statistical mechanics.  Each
\textit{diagonal} element of a reduced statistical density matrix
is a thermal average value having a familiar probabilistic
interpretation similar to (\ref{eq63}).

Expanding (\ref{eq63}), the probability $p_{0,1,1,1,1}(\equiv p)$
of an elementary cavity becomes the difference
\begin{equation}
p=\langle n_1n_2n_3n_4 \rangle - \langle n_0n_1n_2n_3n_4 \rangle .
\label{eq64}
\end{equation}
The representation (\ref{eq64}) is valid for arbitrary lattice gas
interactions (pair, triplet, ...) upon any four-coordinated
regular lattice, i.e., the $d=2$ square and kagom\'e lattices and
the $d=3$ diamond lattice. In Section V, exact solutions for both
RHS correlations in (\ref{eq64}) were obtained along the
coexistence curve  and at the critical point of the
triplet-interaction kagom\'e lattice gas. Consequently, the
probability $p$ in (\ref{eq64}) is likewise determined along the
coexistence curve and at criticality, as well as along the
finite-length critical isofield ($L^*=0$) curve in the disordered
fluid region of $\rho-T$ space. In this setting, the RHS
difference of correlations in (6.4) is seen as the graphical
subtraction of Figure 9 from Figure 8.

The results, viz., $p$ vs $x$, are exhibited in Figure~11 and
Figure~12, the latter being an enlargement of the former in the
range of (reduced) condensation temperatures $0\leq x\leq 1$.  At
absolute zero temperature, the coexistence curve $\rho$~vs~$x$ in
Figure 4a is unity (zero) for the liquid (vapor) branch.  In
either case (fully occupied or completely empty lattice), an
elementary cavity cannot exist, so $p$ vanishes at zero
temperature in Figure 11 (or 12). For increasing temperatures in
Figure~12, the probability $p$ of an elementary cavity along the
liquid branch increasingly exceeds that along the vapor branch,
except at temperatures closely below the critical temperature
($x=1$) where the two branches become identical. Such asymmetric
behavior is physically anticipated since decreasing the particle
number density $\rho_l$ along the liquid branch, beginning from a
fully-occupied ground state liquid phase, creates increasingly
more elementary cavities than increasing $\rho_v$ along the vapor
branch, beginning from a completely empty ground state vapor
phase. Furthermore, the curve $p$~vs~$x$ in the disordered region
($x>1$) of Figure~11 exhibits a rounded \textit{maximum}
$p_{max}=0.0409\cdots$ at $x=\tilde x=3.6\cdots$, and eventually
\textit{vanishes} at $x=5.72739\cdots$ corresponding to $\rho=
\rho_{max}=1$ (fully occupied lattice) in Figure~4a.  The latter
\textit{node} of $p$ along the critical isofield ($L^*=0$)
trajectory is deemed to be a \textit{multiparticle interaction
effect}. In Figure~11, the above maximum value $p_{max}$ is larger
by an order of magnitude than the lower temperature
($x=0.98\cdots$) maximum value along the liquid branch of the
coexistence curve.  A synoptic view of Figures~4a, 11 and 12
enables one to numerically eliminate their common temperature
variable $x$ and obtain the graph $p$~vs $\rho$. \ In particular,
the probability value $p_{max}=0.0409\cdots$ corresponds to a
density value $\rho=\tilde\rho = 0.88\cdots$, and $p$
monotonically increases (decreases) for the density range $ \rho_c
<\rho <\tilde\rho (\tilde\rho < \rho\leq 1)$, where
$\rho_c=0.58931\cdots$ [(3.7a)].

Reflecting upon the traditional case of attractive
nearest-neighbor \textit{pair interactions} in a kagom\'e lattice
gas, the critical isofield ($L^*=0$) solution for $\rho$ in the
disordered region of Figure~4b is a \textit{constant} at all
temperatures, i.e., $1<x<\infty$, $\rho=\frac{1}{2}$
(\textit{half-filled} lattice). The half-filled condition infers
that the probability $p$ of an elementary cavity is
\textit{non-vanishing} along the critical density
($\rho_c=\frac{1}{2}$) line at all temperatures $1<x<\infty$ in
Figure~4b.  For instance, as the (reduced) temperature
$x\rightarrow \infty$, $p\rightarrow
\left(\frac{1}{2}\right)^4-\left(\frac{1}{2}\right)^5=\frac{1}{32}=0.03125$,
having used (\ref{eq64}) with the realization therein that the
average of a product equals the product of the averages at
infinite temperature. The absence of a node in the solution curve
$p$~vs~$x$ for pair interactions at all temperatures $1<x<\infty$
lends credence to the earlier contention that the
finite-temperature node in Figure~11 for triplet interactions is a
multiparticle interaction effect.

\section{Summary and discussion}

Exact results in physics are valuable for a variety of reasons.
Endeavoring to retain only the most essential ingredients of a
physical problem, exact solutions of simple model systems often
provide definite guidance and insights on more realistic and
invariably more mathematically complex systems. Exact results in
tractable models of seemingly different physical systems may alert
researchers to significant common features of these systems and
actually emphasize concepts of universality. In addition to their
own aesthetic appeal, exact results can, of course, serve as
standards against which both approximation methods and approximate
results may be appraised. Also, the underlying mathematical
structures of exactly soluble models in statistical physics are
rich in content and have led to important developments in
mathematics.

Phase diagrams for the condensation of a two-dimensional (d=2)
kagom\'e lattice gas with purely three-particle interactions were
determined previously \cite{barry}. Specifically, the liquid-vapor
phase boundary (chemical potential vs temperature) and the
companion liquid-vapor coexistence curve (density vs temperature)
were obtained. These exact phase diagrams were briefly reviewed in
the present paper to make it self-contained and for later key use
in securing solutions for multiparticle correlations along the
coexistence curve  and at the critical point of the
triplet-interaction kagom\'e lattice gas. Pertinent exact
solutions for the average particle-number density $\langle
n_0\rangle$ and the elementary triplet correlation $\langle
n_0n_1n_2\rangle$ of the lattice gas were obtained  via direct
logarithmic differentiations of the grand canonical partition
function with respect to the (dimensionless) variables $\beta\mu$
and $K_3$, respectively, and then taking an appropriate vanishing
field limit ($L^*\rightarrow 0$). However, to establish solutions
for additional correlations upon the coexistence curve, a
different method was necessary. The method required supplemental
use of the known solutions for $\langle n_0\rangle$ and $\langle
n_0n_1n_2\rangle$, and is briefly discussed below.

A linear algebraic system of correlation identities was generated
having coefficients dependent solely upon the interaction
parameter $K_3$ when evaluated along the coexistence curve. Using
the previously determined solutions for $\langle n_0\rangle$ and
$\langle n_0n_1n_2\rangle$ as a priori information, one succeeded
in finding a closed and linearly independent set of correlation
identities defined upon a spatially compact nine-site cluster
(Figure 6) of the kagom\'e lattice. Employing simple algebraic
techniques, exact solutions are thereby now known for the density
$\langle n_0\rangle$, the nearest-neighbor pair correlation
$\langle n_0n_1\rangle$ and \textit{seventeen} multiparticle
correlations, all along the coexistence curve and at criticality
of the triplet interaction kagom\'e lattice gas. The correlation
solutions were also secured along the continuation of the
curvilinear diameter of the coexistence region into the disordered
fluid region of $\rho-T$ space. To our knowledge, these are the
first examples, certainly away from criticality, of exact
solutions for multiparticle correlations in any planar
lattice-statistical model with multiparticle interactions. Along
the coexistence curve (Figure 4a), the graphs of various
correlation solutions were plotted (Figures 5a, 7a, 8, 9, 10) as
functions of the temperature, and each solution curve exhibited an
asymmetric rounded shape with an infinite slope at the critical
point (solid circle). The degree of asymmetry was associated with
the slope of the curvilinear diameter in the graph, and a negative
slope (Figures 8, 9, 10) was a harbinger for the correlation
solution to have a rounded \textit{minimum} along the continuation
of the curvilinear diameter into the disordered region. The exact
solution curves were compared with the corresponding results in a
conventional half-filled kagom\'e lattice gas having
nearest-neighbor pair interactions, and the solutions for the
multiparticle correlations were applied to cavitation theory in
the condensation of the lattice gas.

The class of correlation identities currently considered has
appeared in the literature \cite{fisher} for almost five decades,
particularly in the context of planar Ising models with
nearest-neighbor pair interactions. In the present paper, however,
the correlation identities were developed within the framework of
the grand canonical ensemble for a kagom\'e lattice gas having
localized three-particle interactions. As reviewed above,
solutions were determined for various localized correlations of
the lattice gas. One further reasons that the lengths of the
vertical ``tie-lines'' connecting the upper and lower branches of
the solution curves vanish with an Ising-type critical exponent
$1/8$, and the curvilinear diameter of each solution curve is
either analytic at the critical point or possesses a weak Ising
energy-type singularity $\epsilon \ln\epsilon$. Supporting
arguments for these asserted behaviors involve the following
particulars. The critical behaviors of the lattice gas model are
embedded in the system of identities at the outset, entering the
initial identities (5.2a,b) via substitution of known solutions
for $\langle n_0\rangle$, $\langle n_0n_1n_2\rangle$ into the
inhomogeneous terms. The remaining identities (5.3), (5.5) and
(5.8) are homogeneous relations among correlations. Solutions for
seventeen correlations are then directly determined by elementary
linear algebraic procedures and cascade orderings within the
system of identities. Aside from well-behaved additive terms
containing solely coefficients (4.6), all solutions can eventually
be represented as linear combinations of  $\langle n_0\rangle$ and
$\langle n_0n_1n_2\rangle$. Hence, one concludes that the familiar
Ising-type critical singularities $\epsilon^{1/8}$, $\epsilon
\ln\epsilon$ inherent in $\langle n_0\rangle$ and $\langle
n_0n_1n_2\rangle$ similarly characterize the singular behaviors in
the subsequent seventeen correlation solutions.

Extensions in these results for the number of correlation
solutions and the shapes and sizes of the $n$-site clusters could
be aided by computer search codes in linear algebra theory. In
these searches, one anticipates linear independence in a system of
identities to be a more elusive algebraic property than closure.
Finally, if one assumes a ``\textit{clustering property}''
(asymptotic separability) for the correlations in the model, one
can then employ the present known solutions for correlations on
smaller length scales to gain information on chosen asymptotically
long-distance-type correlations. A simple example reifies these
concepts. Let $q_1,q_2;$ $r_1,r_2,r_3;$ $s_1,s_2,s_3,s_4,s_5$ be
the lattice sites of a nearest-neighbor bond, an elementary
triangle and a ``bow-tie'' configuration, respectively, where the
$q-$, $r-$, $s-$clusters of sites are mutually separated by
asymptotically large distances. Then, the ``clustering property''
yields the decuplet correlation
\begin{equation}
\langle n_{q_1}n_{q_2}n_{r_1}n_{r_2}n_{r_3}
n_{s_1}n_{s_2}n_{s_3}n_{s_4}n_{s_5}\rangle \sim \langle 01\rangle
\langle 012\rangle\langle 01234\rangle,
\end{equation}
having lastly used lattice symmetry and earlier notations. The
result (7.1) shows that the chosen decuplet correlation
asymptotically equals the \textit{graphical multiplication} of the
solution curves in Figures 5a, 7a and 9.

\newpage

\begin{figure}
\begin{center}
\leavevmode\includegraphics[width=0.3\textheight]{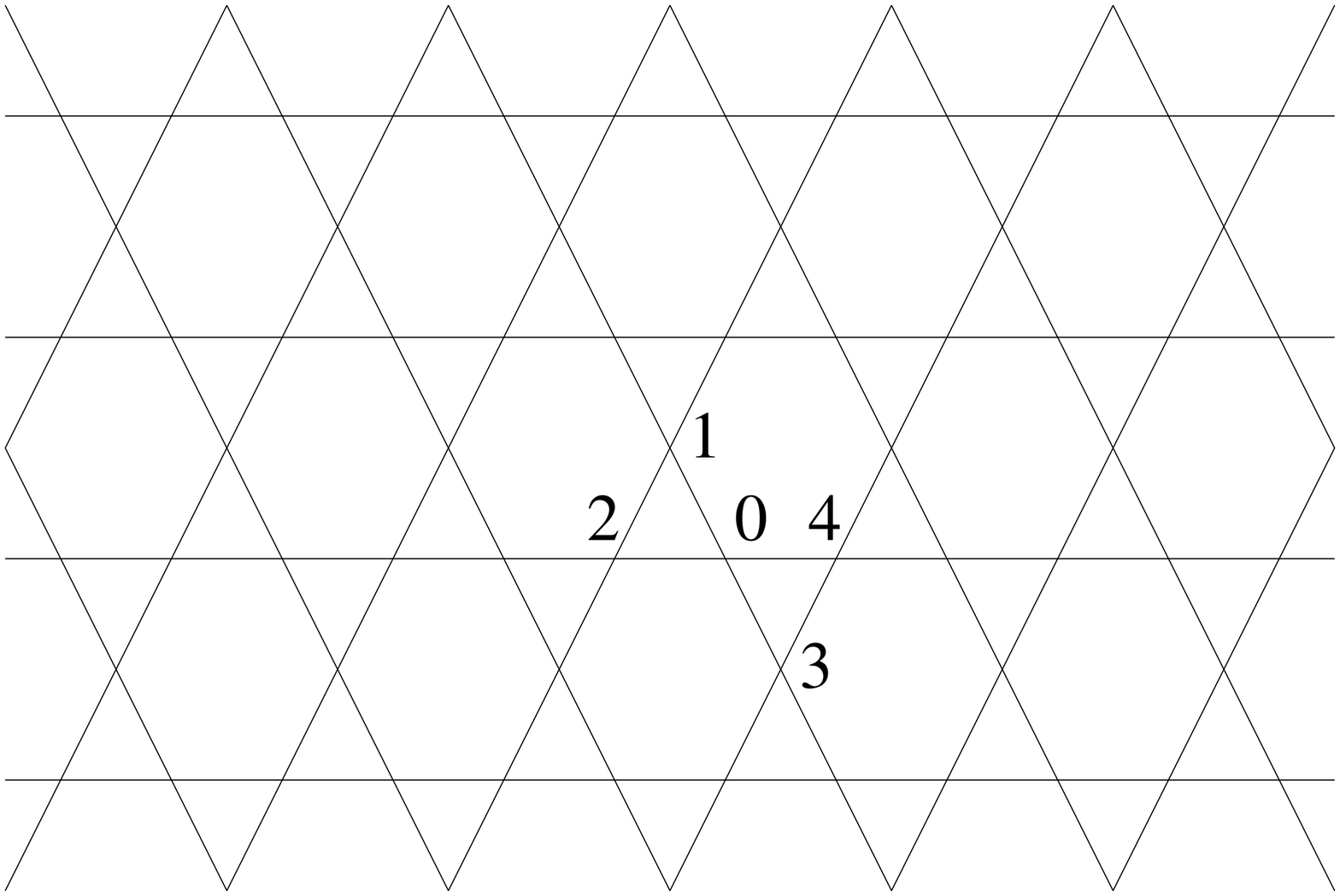}
\end{center}
\caption{The kagom\'e lattice is a two-dimensional periodic array
of equilateral triangles and regular hexagons. The lattice is
regular (all sites equivalent, all bonds equivalent) and has
coordination number 4. The origin site  and its four nearest
neighboring sites are specifically enumerated. Whenever three
atoms of the lattice gas simultaneously occupy the vertices of an
elementary triangle (say sites $0,1,2$), these atoms experience a
triplet interaction with strength parameter $\epsilon_3$. }
\end{figure}

\begin{figure}
\begin{center}
\leavevmode\includegraphics[width=0.3\textheight]{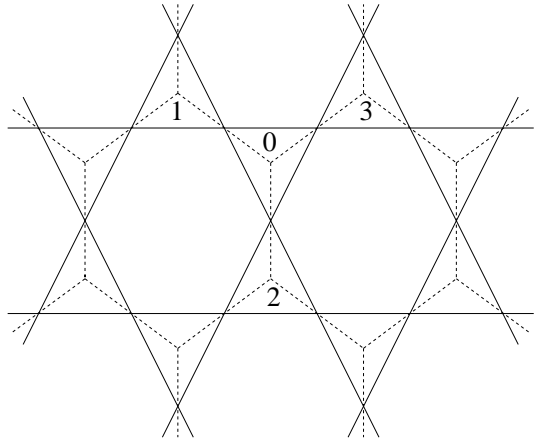}
\end{center}
\caption{A honeycomb lattice (dashed bonds) may be associated with
the kagom\'e lattice (solid bonds). A honeycomb lattice (dashed
bonds) is a two-dimensional periodic array of regular hexagons, is
regular (all sites equivalent, all bonds equivalent) and has
coordination number 3. The origin site and its three
nearest-neighboring sites are specifically enumerated upon the
associated honeycomb lattice. }
\end{figure}

\begin{figure}
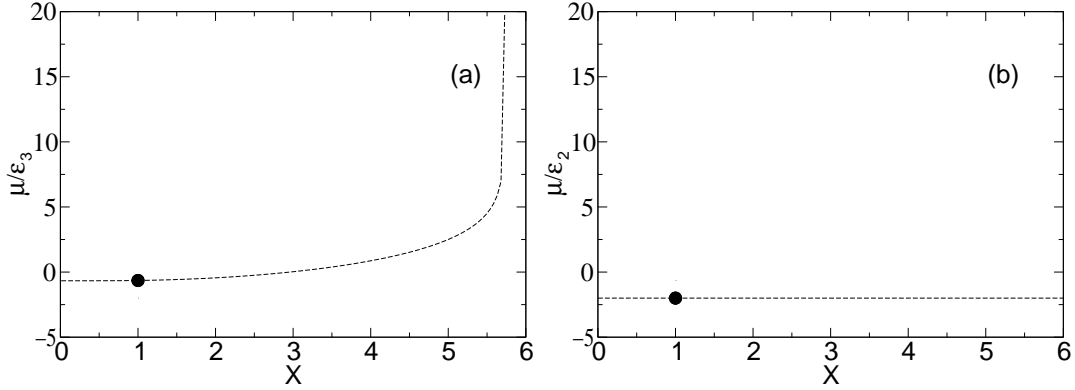

\begin{center}
\includegraphics[clip,angle=270, width=0.3\textheight]{bm-fig03a-revised.eps}
\includegraphics[clip,angle=270, width=0.3\textheight]{bm-fig03b-revised.eps}
\end{center}
\caption{(a) The liquid-vapor phase boundary of the
triplet-interaction kagom\'e lattice gas, with $\mu/\epsilon_3$
being a reduced chemical potential and $x=K_{3c}/K_3 (=T/T_c)$ a
reduced temperature, where $K_{3c}(\equiv\epsilon_3/k_BT_c)
=\ln[(2+\sqrt{3})^3+1] =3.96992\cdots $. The curvilinear phase
boundary begins at zero temperature with $\mu/\epsilon_3=-2/3$,
and ends (analytically) at the critical point (solid circle) whose
coordinates are $x=1$, $\mu/\epsilon_3\equiv \mu_c/\epsilon_3
=-\ln(4\sqrt{3}+6)/\ln(15\sqrt{3}+27) =-0.64469\cdots$. The
continuation of the phase boundary is the dashed curve which
eventually diverges logarithmically at the reduced temperature
$x=K_{3c}/\ln 2=5.72739\cdots$. (b) The liquid-vapor phase
boundary of a conventional kagom\'e lattice gas having
nearest-neighbor pair interactions, with $\mu/\epsilon_2$ being a
reduced chemical potential where $\epsilon_2 >0$ is the strength
parameter of the attractive pair interaction, and $x=K_{2c}/K_2
(=T/T_c)$ being a reduced temperature, where $K_{2c}
=\ln[(3+2\sqrt{3})] =1.86626\cdots $. The phase boundary is
rectilinear, specially, the constant valued expression
$\mu/\epsilon_2 =-2$, $0\le x\le 1$ (solid circle locates the
critical point). The continuation of the phase boundary is the
horizontal dashed line which continues indefinitely. }
\end{figure}

\begin{figure}
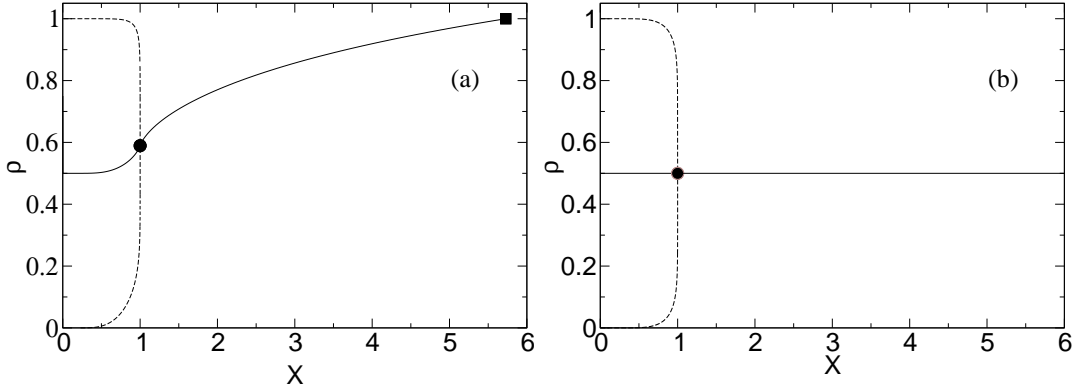

\begin{center}
\includegraphics[clip,angle=270, width=0.3\textheight]{bm-fig04a-revised.eps}
\includegraphics[clip,angle=270, width=0.3\textheight]{bm-fig04b-revised.eps}
\end{center}
\caption{(a) The liquid-vapor coexistence curve of the
triplet-interaction kagom\'e lattice gas, with $\rho$ being the
particle number density and $x=K_{3c}/K_3 (=T/T_c)$ a reduced
temperature, where $K_{3c}(\equiv\epsilon_3/k_BT_c)
=\ln[(2+\sqrt{3})^3+1] =3.96992\cdots $. The curvilinear diameter
of the asymmetric rounded coexistence region is the solid curve
which begins at zero temperature with $\rho=1/2$, and ends at the
critical point (solid circle) whose coordinates are $x=1$,
$\rho\equiv \rho_c =\frac{1}{3}(7/2-\sqrt{3}) =0.58931\cdots$. The
continuation of the curvilinear diameter into the disordered fluid
region is the continuing sigmoidal curve which is monotonically
increasing and concave downward, eventually ending at the point
(solid square) with coordinates  $x=K_{3c}/\ln 2=5.72739\cdots$,
$\rho=\rho_{max}=1$. (b) The liquid-vapor coexistence curve of a
conventional kagom\'e lattice gas having nearest-neighbor pair
interactions, with $\rho$ being the particle number density and
$x=K_{2c}/K_2$ being a reduced temperature where $K_{2c}
=\ln[(3+2\sqrt{3})] =1.86626\cdots $. The coexistence curve is
given by $\rho_{l,v}^{coex}=\frac{1}{2}(1\pm m_s)$, where $m_s$ is
the spontaneous magnetization of a standard kagom\'e Ising model.
The rectilinear diameter of the symmetric rounded coexistence
region is the horizontal solid line $\rho=1/2$, which ends at the
critical point (solid circle) whose coordinates are $x=1$,
$\rho\equiv \rho_c=1/2$. The continuation of the rectilinear
diameter into the disordered region is the horizontal solid line
which continues indefinitely.}
\end{figure}

\begin{figure}
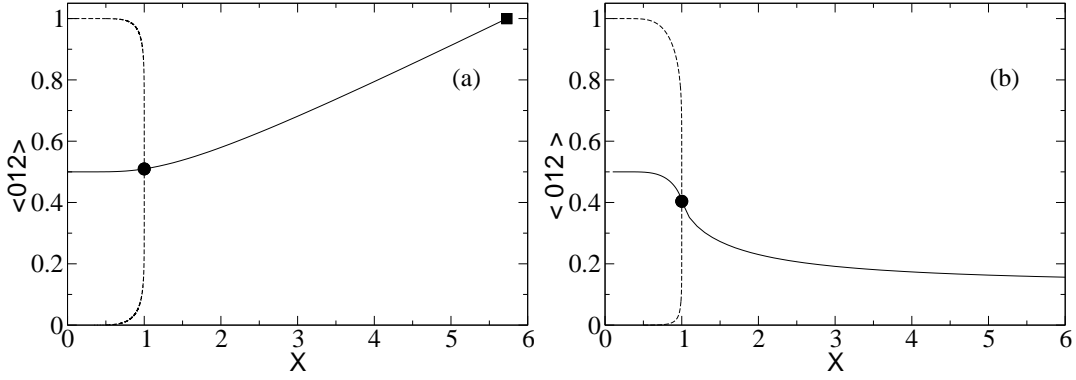

\begin{center}
\includegraphics[clip,angle=270, width=0.3\textheight]{bm-fig05a-revised.eps}
\includegraphics[clip,angle=270, width=0.3\textheight]{bm-fig05b-revised.eps}
\end{center}
\caption{(a) Elementary triplet correlation $\langle 012\rangle$
vs reduced temperature $x=K_{3c}/K_3 (=T/T_c)$, where $K_{3c}
=3.96992\cdots $. The solution (dashed curve) is determined along
the coexistence curve including the critical point (solid circle)
with coordinates $x=1$, $\langle
012\rangle_c=\frac{1}{2}(27+15\sqrt{3})/(26+15\sqrt{3})=0.50961\cdots$.
The solution is also secured along the continuation of the
curvilinear diameter (solid curve) into the disordered region,
viz., the monotonically increasing and concave-upward curve,
eventually ending at the point (solid square) with coordinates
$x=K_{3c}/\ln 2=5.72739\cdots$, $\langle 012\rangle=1$. (b)
Conventional pair-interaction kagom\'e lattice gas. Elementary
triplet correlation $\langle 012\rangle$ vs reduced temperature
$x=K_{2c}/K_2$ where $K_{2c} =1.86626\cdots $. The solution
(dashed curve) is determined along the coexistence curve including
the critical point (solid circle) with coordinates $x=1$, $\langle
012\rangle_c=\frac{1}{16}(3+2\sqrt{3})=0.40400\cdots$. The
solution is also secured along the continuation of the curvilinear
diameter (solid curve) into the disordered region, viz., the
continuing sigmoidal curve which is monotonically decreasing and
concave upward, eventually approaching $(\frac{1}{2})^3=0.125$ at
infinite temperature.}
\end{figure}

\begin{figure}
\begin{center}
\leavevmode\includegraphics[width=0.25\textheight]{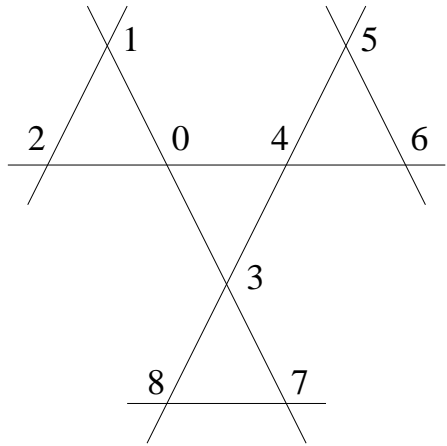}
\end{center}
\caption{Nine sites of the kagom\'e lattice are specifically
enumerated for selected applications of the theory.}
\end{figure}

\begin{figure}
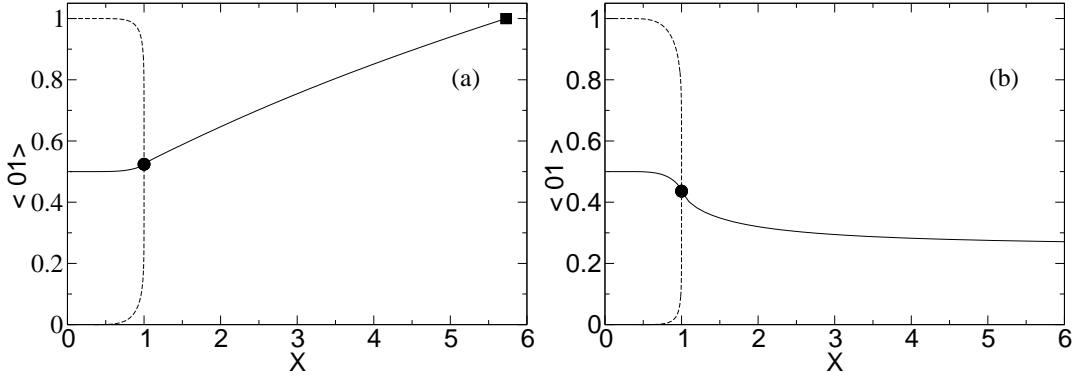

\begin{center}
\includegraphics[clip,angle=270, width=0.3\textheight]{bm-fig07a-revised.eps}
\includegraphics[clip,angle=270, width=0.3\textheight]{bm-fig07b-revised.eps}
\end{center}
\caption{(a) Nearest neighbor pair correlation $\langle 01\rangle$
vs reduced temperature $x=K_{3c}/K_3 (=T/T_c)$ where $K_{3c}
=\ln[(2+\sqrt{3})^3+1] =3.96992\cdots $. The solution (dashed
curve) is determined along the coexistence curve including the
critical point (solid circle) with coordinates $x=1$, $\langle
01\rangle_c=0.5239\cdots$. The solution is also secured along the
continuation of the curvilinear diameter (solid curve) into the
disordered region, viz., the continuing sigmoidal curve which is
monotonically increasing and slightly concave downward, eventually
ending at the point (solid square) with coordinates $x=K_{3c}/\ln
2=5.72739\cdots$, $\langle 01\rangle=1$. (b) Conventional
pair-interaction kagom\'e lattice gas. Nearest-neighbor pair
correlation $\langle 01\rangle$ vs reduced temperature
$x=K_{2c}/K_2 (=T/T_c)$ where $K_{2c} =\ln[(3+2\sqrt{3})]
=1.86626\cdots $. The solution (dashed curve) is determined along
the coexistence curve including the critical point (solid circle)
with coordinates $x=1$, $\langle
01\rangle_c=\frac{1}{12}(\frac{7}{2}+\sqrt{3}) =0.43600\cdots$.
The solution is also secured along the continuation of the
curvilinear diameter (solid curve) into the disordered region,
viz., the continuing sigmoidal curve which is monotonically
decreasing and concave upward, eventually approaching
$(\frac{1}{2})^2=0.25$ at infinite temperature.}
\end{figure}

\begin{figure}
\begin{center}
\includegraphics[angle=270, width=0.3\textheight]{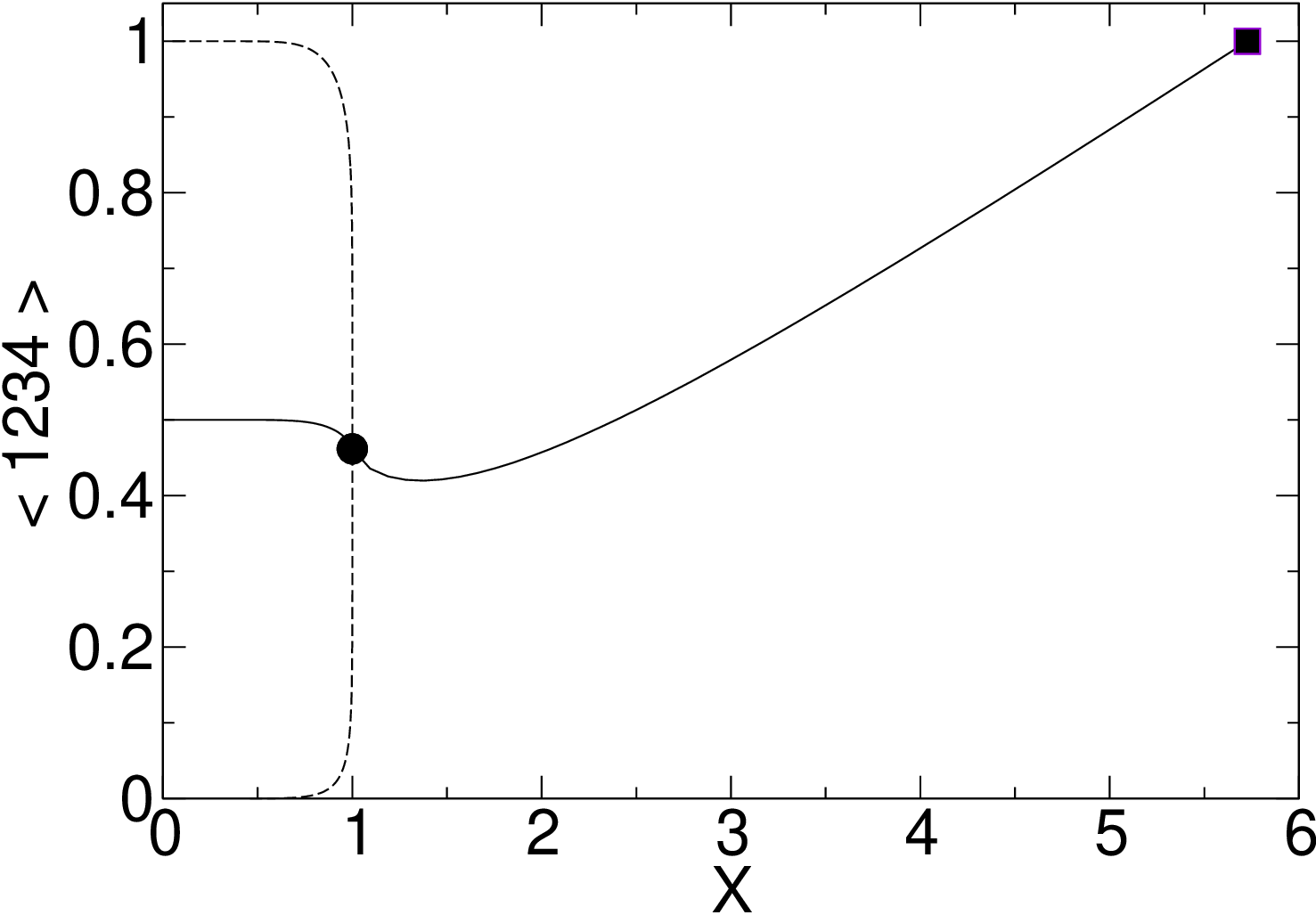}
\end{center}
\caption{Quartet correlation $\langle 1234\rangle$ vs reduced
temperature $x=K_{3c}/K_3 (=T/T_c)$ where $K_{3c} =3.96992\cdots
$. The solution (dashed curve) is determined along the coexistence
curve including the critical point (solid circle) with coordinates
$x=1$, $\langle 1234\rangle_c=0.4617\cdots$. The solution is also
secured along the continuation of the curvilinear diameter (solid
curve) into the disordered region, viz., a concave upward curve
having an asymmetric rounded minimum before monotonically
increasing towards an end point (solid square) with coordinates
$x=K_{3c}/\ln 2=5.72739\cdots$, $\langle 1234\rangle=1$. }
\end{figure}

\begin{figure}
\begin{center}
\includegraphics[angle=270, width=0.3\textheight]{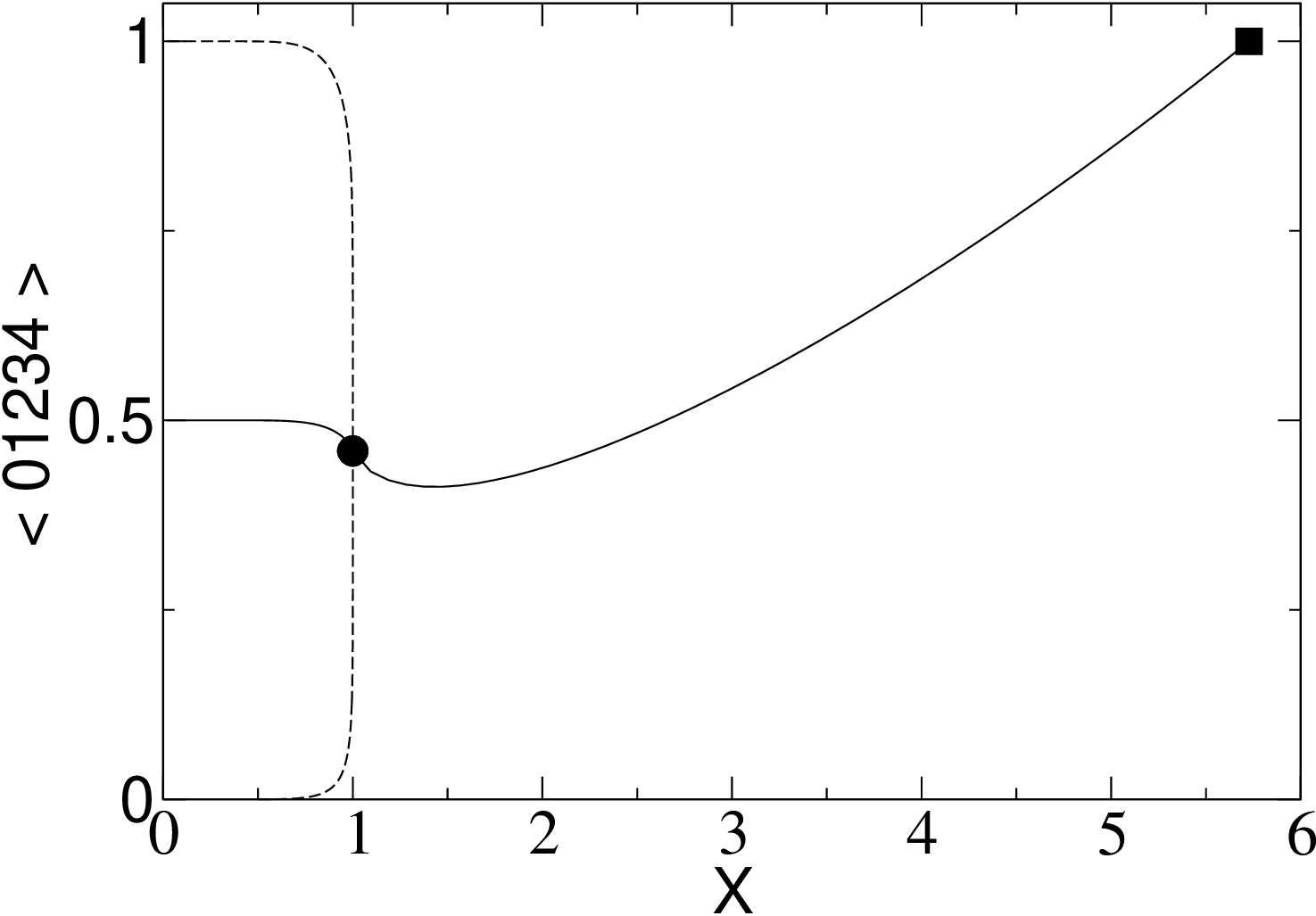}
\end{center}
\caption{Quintet correlation $\langle 01234\rangle$ vs reduced
temperature $x=K_{3c}/K_3 (=T/T_c)$ where $K_{3c} =3.96992\cdots
$. The solution (dashed curve) is determined along the coexistence
curve including the critical point (solid circle) with coordinates
$x=1$, $\langle 01234\rangle_c=0.4596\cdots$. The solution is also
secured along the continuation of the curvilinear diameter (solid
curve) into the disordered region, viz., a concave upward curve
having an asymmetric rounded minimum before monotonically
increasing towards an end point (solid square) with coordinates
$x=K_{3c}/\ln 2=5.72739\cdots$, $\langle 01234\rangle=1$.}
\end{figure}

\begin{figure}
\begin{center}
\includegraphics[angle=270, width=0.3\textheight]{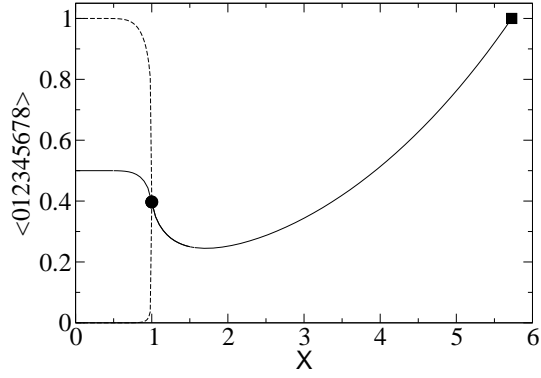}
\end{center}
\caption{Nonuplet correlation $\langle 012345678\rangle$ vs
reduced temperature $x=K_{3c}/K_3 (=T/T_c)$ where $K_{3c}
=3.96992\cdots $. The solution (dashed curve) is determined along
the coexistence curve including the critical point (solid circle)
with coordinates $x=1$, $\langle 012345678\rangle_c=0.3970\cdots$.
The solution is also secured along the continuation of the
curvilinear diameter (solid curve) into the disordered region,
viz., a concave-upward curve having an asymmetric rounded minimum
before monotonically increasing towards an end point (solid
square) with coordinates $x=K_{3c}/\ln 2=5.72739\cdots$, $\langle
012345678\rangle=1$.}
\end{figure}

\begin{figure}
\begin{center}
\leavevmode
\includegraphics[angle=270, width=0.3\textheight]{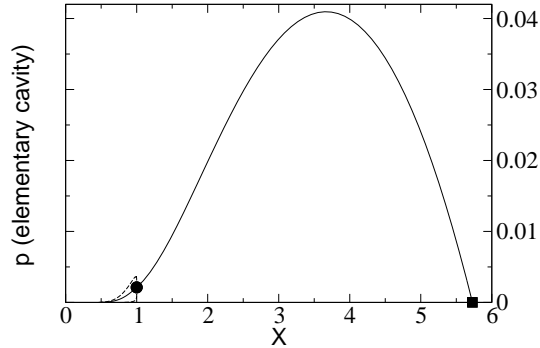}
\end{center}
\caption{Probability $p$ of an elementary cavity vs reduced
temperature $x=K_{3c}/K_3 (=T/T_c)$ where $K_{3c} =3.96992\cdots
$. The solution (dashed curve) is determined along the coexistence
curve including the critical point (solid circle) with coordinates
$x=1$, $p_c=\langle 1234\rangle_c-\langle 01234\rangle_c
=0.0021\cdots$. The solution is also secured along the
continuation of the curvilinear diameter (solid curve) into the
disordered region, viz., a concave-downward curve with a symmetric
rounded maximum located at $\tilde{x}=3.6\cdots$,
$p=0.0409\cdots$, before monotonically decreasing towards a nodal
end point (solid square) at $x=K_{3c}/\ln 2=5.72739\cdots$ (fully
occupied lattice).}
\end{figure}

\begin{figure}
\begin{center}
\leavevmode
\includegraphics[angle=270, width=0.3\textheight]{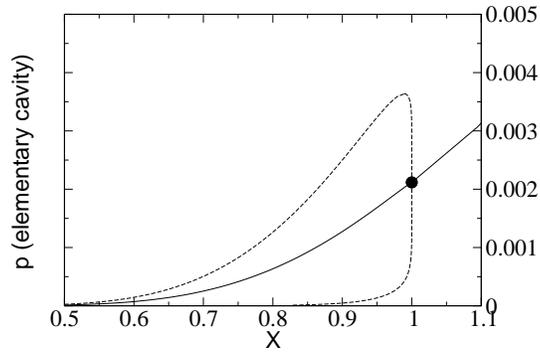}
\end{center}
\caption{Enlargement of Fig 11 in the range of (reduced)
condensation temperatures $0\le x\le 1$. Along the upper liquid
branch $\rho_l$ of the coexistence curve, the probability $p$ of
an elementary cavity attains an asymmetric rounded maximum located
at $x=0.98\cdots$, $p=0.0036\cdots$. The solid circle is the
critical point with coordinates $x=1$, $p_c=0.0021\cdots$.}
\end{figure}

\end{document}